\documentclass[12pt]{article}
\usepackage{amsmath,amssymb,epsfig}

\usepackage{color}
\input{colordvi.tex}

\renewcommand{\thefootnote}{\fnsymbol{footnote}}

\makeatletter

\renewcommand\section{\@startsection {section}{1}{\z@}%
                                   {-3.5ex \@plus -1ex \@minus -.2ex}%
                                   {2.3ex \@plus.2ex}%
                                   {\normalfont\large\bfseries}}

\renewcommand\subsection{\@startsection{subsection}{2}{\z@}%
                                     {-3.25ex\@plus -1ex \@minus -.2ex}%
                                     {1.5ex \@plus .2ex}%
                                     {\normalfont\normalsize\bfseries}}

\newcount\hour \newcount\minute
\hour=\time \divide \hour by 60
\minute=\time
\def\now{%
\ifnum \hour<13
  \ifnum \hour=0 \advance \hour by 12 \number\hour:\else \number\hour:\fi%
     \ifnum \minute<10 0\fi%
     \number\minute%
\ A.M.%
\else \advance \hour by -12 \number\hour:%
  \ifnum \minute<10 0\fi%
  \number\minute%
  \ P.M.%
\fi%
}

\makeatother

\begin{document}

\baselineskip=18pt  
\numberwithin{equation}{section}  
\allowdisplaybreaks  



%
%


\thispagestyle{empty}

\vspace*{-2cm}
\begin{flushright}
\end{flushright}

\begin{flushright}
YITP-10-10\\ 
IPMU-10-0188
\end{flushright}

\begin{center}

\vspace{2.5cm}

{\bf\Large Comments on Gaugino Mass and Landscape of Vacua}
\\

\vspace*{1.5cm}
{\bf
Yuichiro Nakai$^{1}$\footnote{e-mail:
{\tt ynakai@yukawa.kyoto-u.ac.jp}} and Yutaka Ookouchi$^{2,3}$}\footnote{e-mail:
{\tt yutaka.ookouchi@ipmu.jp}} \\
\vspace*{0.5cm}

$^{1}${\it {Yukawa Institute for Theoretical Physics, Kyoto University,
   Kyoto 606-8502, Japan}}
\vspace{0.1cm}

$^{2}${\it Perimeter Institute for Theoretical Physics, Ontario N2L2Y5, Canada}\\
\vspace{0.1cm}

$^{3}${\it Institute for the Physics and Mathematics of the Universe (IPMU), \\
University of Tokyo, Chiba, 277-8583, Japan}\\
\vspace{0.1cm}

\end{center}

\vspace{1cm} \centerline{\bf Abstract} \vspace*{0.5cm}

In direct gauge mediation, gaugino masses often vanish at the leading order of supersymmetry breaking.
Recently, this phenomenon is understood in connection with the global structure of vacua in O'Raifeartaigh-type models.
In this note, we further explore a connection between gaugino masses and the landscape of vacua in more general situations, focusing on a few examples which demonstrate our idea.
In particular, we present a calculable model with non-vanishing leading order gaugino masses on the lowest energy vacuum.

\newpage
\setcounter{page}{1} 
\setcounter{footnote}{0}
\renewcommand{\thefootnote}{\arabic{footnote}}





\section{Introduction}

Gauge mediation \cite{GaugeMediation} is a highly predictive and an attractive way of transmitting supersymmetry breaking of a hidden sector to the supersymmetric standard model.  Many gauge mediation models are already known (See \cite{Giudice:1998bp,Kitano:2010fa} for reviews). Among them, direct gauge mediation is the model that the flavor symmetries in a hidden SUSY breaking sector are weakly gauged and identified as the standard model gauge symmetries. In most controllable direct gauge mediation models, the scale of supersymmetry breaking $\sqrt{F}$ (the square-root of the F-term of a SUSY breaking field) is much smaller than the messenger scale $M$. In such models, the leading contributions to the standard model gaugino masses in an $F/M^2$ expansion can be calculated by a powerful tool, analytic continuation into superspace \cite{GR,AGLR}. However, it often happens in direct gauge mediation models that the leading contributions to gaugino masses vanish even if R-symmetry is broken, firstly emphasized in \cite{IzawaTobe}.\footnote{Anomalously small gaugino mass problem also occur in another class of gauge mediation models such as semi-direct gauge mediation \cite{Semidirect}. In this class, sizable gaugino masses can be generated by introducing a strongly coupled messenger sector \cite{IIN1}.} Since there are no such cancellations for scalar masses, this implies that the gauginos are much lighter than the scalars.\footnote{We can suppress the scalar masses by exploiting a superconformal dynamics \cite{HO}.} Then, we cannot obtain order 1 TeV gaugino masses and scalar masses at the same time, which causes the standard model hierarchy problem again. Furthermore, recent experimental data severely constrain such a model with the light gauginos \cite{Yonekura}. 
 Thus, it is an interesting challenge to generate sizable gaugino masses in direct gauge mediation. The authors of \cite{KOO} avoided this issue by exploiting a higher energy metastable state (This possibility was recently emphasized in \cite{GKK,Abel:2009ze}). However, the reason why an uplifted vacuum generates sizable gaugino masses has remained mysterious.

Recently, Komargodski and Shih (KS) shed light on this curious feature of direct gauge mediation and clarified that the pseudomoduli space cannot be locally stable everywhere in order to generate sizable gaugino masses \cite{KS}. It is interesting that anomalously small gaugino masses are closely related to the global structure of vacua. It is worth exploring this connection further in more general situations.

In this paper, we survey the connection not only in direct gauge mediation but also in minimal gauge mediation \cite{Dine:1994vc}. We exploit non-canonical K\"ahler potential of messengers and provide a way to generate the leading order gaugino masses while keeping all messenger directions stable everywhere. Furthermore, we construct a minimal-type gauge mediation model with non-vanishing leading order gaugino masses on the lowest energy vacuum.

The rest of the paper is organized as follows. In section $2$, we will review a property in O'Raifeartaigh-type models  with canonical K\"ahler potential and a connection between the global structure of vacua and the leading order gaugino masses. In section $3$, we will show how non-canonical K\"ahler potential of messenger fields alters the stability of messenger directions at a point of the pseudomoduli space. We also provide an explicit model which has non-zero leading order gaugino masses and a pseudomoduli space that is locally stable everywhere. In section $4$, we present a calculable model with an additional $U(1)$ gauge interaction which generates the leading order gaugino masses on the lowest energy vacuum.

\section{Review of the KS theorem}

First, we will briefly describe the existence of a pseudomoduli space discussed in \cite{KS,Ray} by exploiting a simple example. Let us consider the following Wess-Zumino model with canonical K\"ahler potential presented in \cite{Cheung:2007es},
\begin{equation}
W= \lambda X(\phi_1\tilde{\phi_1}+\phi_2\tilde{\phi_2}) + m\phi_1\tilde{\phi_2} + fX,
\end{equation}
where $X$ is a SUSY breaking field and $\phi_1, \tilde{\phi_1}, \phi_2, \tilde{\phi_2}$ are vector-like pairs of messenger fields charged under the standard model gauge symmetries. We can choose parameters $\lambda, m, f$ as real without loss of generality. On the metastable SUSY breaking vacuum, $\langle \phi_1 \rangle = \langle \tilde{\phi_1} \rangle = \langle \phi_2 \rangle = \langle \tilde{\phi_2} \rangle = 0$ and $\langle X \rangle$ is the pseudomoduli space of vacua. That is, it takes an arbitrary value at the classical level, but takes a definite value by quantum effects. In this model, as discussed in \cite{Cheung:2007es}, the vev $\langle X \rangle$ takes a non-zero value, so R-symmetry is spontaneously broken. As this example, every O'Raifeartaigh-type model with canonical K\"ahler potential has a pseudomoduli space on every SUSY breaking vacuum.

Next, we will see a connection between the global structure of vacua and the leading order gaugino masses by dealing with the same example as above. The leading order gaugino masses are calculated by analytic continuation into superspace technique such as,
\begin{equation}
m_{\tilde{g}} \sim f \frac{\partial}{\partial X} \log \det \mathcal{M}_F,
\end{equation}
where $\mathcal{M}_F$ is the fermion mass matrix of messengers. In this model, the determinant of the fermion mass matrix is given by $\det \mathcal{M}_F = \lambda^2 X^2$ and has an $X$ dependence, so gaugino masses are generated at the leading order of the SUSY breaking $f$,
\begin{equation}
m_{\tilde{g}} \sim \frac{f}{\langle X \rangle}.
\end{equation}
On the other hand, in this model, there is a zero eigenvalue in the fermion mass matrix at $\langle X \rangle = 0$. Here, the eigenvalues of the scalar mass matrix are $\left( m^2 \pm \sqrt{m^4 + 4 \lambda^2 f^2} \right)/2$, so there is a tachyonic direction. As this example, there is always at least one tachyonic direction on the pseudomoduli space of vacua when gaugino masses are generated at the leading order. Otherwise, the leading order gaugino masses vanish. For general arguments, see the original paper \cite{KS}.

\section{General cases}

As we reviewed in the previous section, in every renormalizable O'Raifeartaigh-type
model, the pseudomoduli space cannot be stable everywhere to generate gaugino masses.
However, a renormalizable model is not always a good description of dynamical
SUSY breaking at low-energy. In many SUSY breaking models, correction terms in K\"ahler potential
are not negligible. In this section, we emphasize that such terms
affect crucially the connection between gaugino masses and the landscape of vacua. 

\subsection{Stability of messenger directions}

Let us start with a general argument for the stability of
messenger directions. Suppose we have a superpotential interaction, 
\begin{equation}
W={\cal M}_F (X)_{ab}\tilde{\phi}^a\phi^b+f(X),
\end{equation}
where $X$ is a chiral superfield which is responsible of SUSY breaking and $\phi, \tilde{\phi}$ are messengers. The lower indices of the messenger mass matrix ${\cal M}_F$ denote the derivatives with respect to messenger fields. When we turn on a generic non-canonical K\"ahler potential, $X$ direction is not
necessarily pseudo-flat as discussed in \cite{Ray}.\footnote{Also, D-term SUSY breaking models
do not always have pseudomoduli spaces.} Nevertheless, in order to focus on the stability of messenger directions at a point of the pseudomoduli space like \cite{KS}, we can keep a flat direction by imposing the following condition on the metric
\cite{AM},
\begin{equation}
\partial_X g^{X\bar{X}}\big|_{0} =0, \label{cond1}
\end{equation}
where $|_0$ denotes $\langle \phi^a  \rangle=\langle \tilde{\phi}^a  \rangle =0$. It is easy to check that the scalar potential with this condition keeps $X$ direction flat. 

In this setup, the boson mass-squared matrix of the messengers is given by 
\begin{equation}
{\cal M}_B^2=\left(
  \begin{array}{cc}
  (\mathcal{M}_F^{\ast}\mathcal{M}_F)_{a\bar{b}} - {\cal A}_{a\bar{b}} &
\mathcal{F}^{\ast}_{ab} \\
  \mathcal{F}_{\bar{a}\bar{b}} & (\mathcal{M}_F\mathcal{M}_F^{\ast})_{\bar{a} {b}}-
{\cal A}_{\bar{a}b}
  \end{array}
 \right).
\end{equation}
Here,
\begin{equation}
{\cal F}_{ab}= F_X^* (\partial_{X} {{\cal M}_F})_{ab},\quad {\cal
A}_{a\bar{b}}= {R_{a \bar{b}}}^{X\bar{X}} |F_X|^2,
\end{equation}
where ${R_{a\bar{b}}}^{X\bar{X}}$ are components of the Riemann tensor. We simply assumed
$g_{X\bar{X}}=1$ at $\tilde{\phi}^a=\phi^b=0$. Suppose $v_a$ is a unit vector satisfying $(\mathcal{M}_F)_{ab} v_b = 0$. Then, a bosonic
mode corresponding to this direction has a mass, 
\begin{equation}
\begin{split}
 &\left(
 v^\dagger~ v^T
 \right)
 {\cal M}_B^2
 \left(
  \begin{array}{c}
   v \\
   v^{\ast}
  \end{array}
 \right)= v^{T}\mathcal{F}v -v^\dagger {\cal A} v + c.c.
 \end{split}
\end{equation}
If ${\cal A}v=0$ or simply if ${\cal A}=0$, then the bosonic mode must be massless in order to have a consistent vacuum, or we have to allow the vacuum to have a tachyonic direction. However, in general, this does not true. As we will demonstrate below, one can easily lift a tachyonic direction and make the pseudomoduli space stable everywhere by using the contribution from the non-canonical part of K\"ahler potential ${\cal A}$.

\subsection{A model with non-canonical K\"ahler potential}

In the rest of this section, we will try to construct the model which has non-zero leading order gaugino masses and a pseudomoduli space that is locally stable everywhere. Let us consider the model discussed in section $2$ with non-canonical K\"ahler potential, that is, its superpotential is given by
\begin{equation}
W= \lambda X(\phi_1\tilde{\phi_1}+\phi_2\tilde{\phi_2}) + m\phi_1\tilde{\phi_2} + fX.
\end{equation}
Notation is explained in the previous section. This model with canonical K\"ahler potential has a tachyonic direction around $\langle X \rangle = 0$, so we will try to lift this direction by introducing non-canonical K\"ahler potential,\footnote{Although we consider a UV-insensitive model and our result does not depend on the way of a UV completion, the authors tried to obtain this kind of non-renormalizable theory from a renormalizable one by integrating out some heavy modes at the tree-level. However, it seems to be difficult to find the UV completion with massive vector bosons of additional gauge groups keeping the flat direction. Integrating out massive chiral superfields also does not seem to be a workable strategy. It might be viable approach to consider the quantum effects of additional fields.}
\begin{equation}
K=|X|^2+\left(1+ {|X|^2\over M^2} \right) \left(|\phi_1|^2+ |\tilde{\phi}_2|^2 \right)+\left(1- {|X|^2\over M^2} \right) \left(|\tilde{\phi}_1|^2+|\phi_2|^2 \right), \label{noncanonical}
\end{equation}
where $M$ is a large cut-off scale of the theory and we have required vanishing messenger mass supertrace so that our model is UV insensitive \cite{Poppitz,EGGM}. Since the above K\"ahler potential satisfies the condition \eqref{cond1} given in the previous subsection, the pseudo flat-direction of $X$ is kept. There is a zero eigenvalue in the fermion mass matrix at $\langle X \rangle = 0$, and here the eigenvalues of the boson mass-squared matrix of messengers are 
\begin{equation}
\frac{1}{2}\left(\,{m}^{2} \pm \,\sqrt {{m}^{4}+4\,{\lambda}^{2}{f}^{2}-4\,(f/M)^{2}{m}^{2}+4\,(f/M)^{4}} \right).
\end{equation}
We can impose a condition between parameters of the model such as ${\lambda}^{2}{f}^{2}-\,(f/M)^{2}{m}^{2}+\,(f/M)^{4}<0$ so as not to have any tachyonic direction. Since non-canonical K\"ahler potential of messengers does not contribute to the gaugino mass at the leading order, known as gaugino screening \cite{AGLR}, the leading order gaugino mass is given in the same fashion as the case with canonical K\"ahler potential,\footnote{Unlike the gaugino mass, non-canonical K\"ahler potential of messengers contributes to the scalar mass at the leading order.}
\begin{equation}
m_{\tilde{g}} \sim \frac{f}{\langle X \rangle}.
\end{equation}
Here, the expectation value of $X$ can be estimated by stabilizing the one-loop effective potential. The Coleman-Weinberg potential in this kind of models has been calculated in \cite{AM}, which claims that $X$ does not stabilize at the origin and so R-symmetry is spontaneously broken even in the case with non-canonical K\"ahler potential. Therefore, we can obtain non-zero leading order gaugino masses in the model with a pseudomoduli space that is locally stable everywhere.

While we have considered a model with a pseudo-flat direction, as we have seen in the previous subsection, the existence of pseudomoduli is not guaranteed in models with non-canonical K\"ahler potential. In the next section, we will see the relation between this kind of models and the landscape of vacua.

\section{Sizable gaugino mass on the global minimum}

When we consider the case where there is no pseudomoduli space, it becomes unclear how we can generalize the statement of the KS theorem. We are interested in a connection between the leading order gaugino masses and metastability of the vacuum. Then, we will try to solve the question whether we can obtain non-vanishing gaugino masses on the global minimum or not. The answer is yes. In \cite{Nomura:1997ur}, the authors obtained non-vanishing leading order gaugino masses on the global minimum (See also \cite{Ibe:2010jb} for a related work on this avenue realizing the ultra-light gravitino mass). However, they used a dynamical SUSY breaking model and the resulting model is incalculable. Then, for our current purpose, we do not need to focus on dynamical SUSY breaking, so we can take our familiar O'Raifeartaigh-type model.

The explicit model of the SUSY breaking sector is a $U(1)$ gauge theory\footnote{The $U(1)$ gauge coupling becomes strong at high energy. As a logical possibility, there may be a state which has lower energy than the state we now consider by non-perturbative effects. Then, for more rigorous statement, we try to construct a gauge mediation model with the global minimum at least perturbatively.} whose superpotential is given by\footnote{This model was studied in \cite{Dine:2006xt}.}
\begin{equation}
W = X_0(f+\lambda \varphi_1\varphi_2)+ m(X_1\varphi_1+X_2\varphi_2).
\end{equation}
The $U(1)$ charge assignments of $X_0$, $X_1$, $X_2$, $\varphi_1$ and $\varphi_2$ are $0$, $-1$, $1$, $1$ and $-1$ respectively.
We call this $U(1)$ gauge interaction as the messenger gauge interaction.
We can take all couplings, $\lambda, m, f$ as real without loss of generality and assume $f \ll m^2$. On the SUSY breaking vacuum, $\langle X_1 \rangle = \langle X_2 \rangle = \langle \varphi_1 \rangle = \langle \varphi_2 \rangle = 0$ and $X_0$ has a non-zero F-term.

Next, consider the messenger sector. The simplest possibility for our purpose would be the following, 
\begin{equation}
W_{mess}= y_q S q\tilde{q} +y_ES E\tilde{E} + {\kappa \over 3} S^3,
\end{equation}
where $q$ and $\tilde{q}$ are messengers charged under the standard model gauge symmetries and $S, E, \tilde{E}$ are the standard model gauge singlets. Only $E, \tilde{E}$ have charges $1,-1$ under the messenger $U(1)$ gauge interaction. We also take couplings $y_q, y_E, \kappa$ as real. When we integrate out the SUSY breaking sector, two-loop correction generates positive scalar masses for fields $E$ and $\tilde{E}$ like usual gauge mediation, which is given by
\begin{equation}
m_E^2 = m_{\tilde{E}}^2 \sim \left({g_{mess}^2\over 16\pi^2} \right)^2 \left({\lambda f \over m} \right)^2,
\end{equation}
where $g_{mess}$ is the coupling of the messenger gauge interaction. As pointed out in \cite{Nomura:1997ur}, these positive scalar masses generate negative mass squared by one-loop effects of $E$ and $\tilde{E}$ such as
\begin{equation}
-m_S^2 \simeq {4\over 16\pi^2}y^2_E m_E^2 \ln {\Lambda \over m_E},
\label{Smass}
\end{equation}
where $\Lambda$ is the cut-off scale and we assume $y_E \lesssim 1$ so that $m_E^2 \gg |m_S^2|$ is satisfied.
Then, the effective scalar potential of the messenger sector including these corrections is given by
\begin{eqnarray}
V_{mess}&=&\big|{y_E S \tilde{E}} \big|^2+\big|y_E{SE} \big|^2+\big|{y_q S \tilde{q}} \big|^2+\big|y_q{Sq} \big|^2+\big|{y_E E\tilde{E}+y_q q\tilde{q}+\kappa S^2} \big|^2\nonumber \\
&&+m_E^2 |E|^2 +m_E^2 |\tilde{E}|^2+m_S^2 |S|^2. 
\end{eqnarray}
This potential is minimized at
\begin{equation}
\langle |S|^2 \rangle ={|m_S^2| \over 2\kappa^2},\ \  \langle q \rangle =\langle \tilde{q} \rangle =\langle E \rangle =\langle \tilde{E} \rangle =0.
\end{equation}
Note that the expectation value of the SUSY breaking field $S$ is uniquely determined and there is no pseudomoduli space in the messenger sector. The contribution to the vacuum energy is given by
\begin{equation}
V_0=-{m_S^4 \over 4 \kappa^2}.
\end{equation}
This vacuum is the global minimum in certain parameter range. The standard model gaugino mass can be calculated as
\begin{equation}
m_{\tilde{g}} \sim  {\langle |F_S| \rangle \over \langle S \rangle } = \frac{|m_S|}{\sqrt{2}}.
\end{equation}
Therefore, we obtain the leading order gaugino mass on the global minimum of the potential, unlike direct gauge mediation without additional gauge interactions.\footnote{If we express the gaugino mass in terms of the original SUSY breaking $f$, the gaugino mass squared comes from a three-loop effect. However, the scalar mass is also suppressed and both masses are almost the same size as in minimal gauge mediation \cite{Dine:1994vc}.}

While constructing a model with no phenomenological problem is not the purpose of this paper, we finally comment on some points. Although the messenger $U(1)$ gauge boson is massless, it may cause no problem, since the standard model quarks and leptons do not have the messenger $U(1)$ charges. However, if we want to avoid the existence of additional massless gauge bosons, we can higgs the gauge symmetry by adding the following superpotential,
\begin{equation}
W = hT(\Psi\tilde{\Psi}-v^2),
\end{equation}
where $\Psi, \tilde{\Psi}$ are a vector-like pair of chiral superfields charged under the messenger $U(1)$ gauge interaction and $T$ is a Lagrange multiplier field. $h$ is a coupling constant. This term does not restore SUSY, so our argument in this section remains true even if we add this term to the above model. Furthermore, in the SUSY breaking sector, there is a parity given by
\begin{equation}
X_1 \leftrightarrow X_2, \,\,\, \varphi_1 \leftrightarrow \varphi_2, \,\,\, 
E \leftrightarrow \tilde{E}, \,\,\, V \leftrightarrow -V,
\end{equation}
and so the problematic FI term of the messenger $U(1)$ gauge field is forbidden.

\bigskip

\appendix

\setcounter{equation}{0}
\renewcommand{\theequation}{A.\arabic{equation}}
\appendix

\section*{Acknowledgments}

We would like to thank M. Ibe, Z. Komargodski, D. Shih and T. Yanagida for useful discussions and comments. YO's research is supported by World Premier International Research Center Initiative (WPI Initiative), MEXT, Japan. 
YN is grateful to the Perimeter Institute for Theoretical Physics for their hospitality during the course of this work. YN and YO would like to thank the Institute for advanced study for their hospitality. YO's research at the Perimeter Institute for Theoretical Physics is supported in part by the Government of Canada through NSERC and by the Province of Ontario through MRI. 


\bigskip

\appendix

\setcounter{equation}{0}
\renewcommand{\theequation}{A.\arabic{equation}}
\appendix

%
%

\end{document}